\documentclass[12pt,a4paper]{article}

\usepackage{fullpage}
\usepackage[utf8]{inputenc}
\usepackage{amssymb}
\usepackage{amsmath}
\usepackage{amsthm}
\usepackage[round]{natbib}
\usepackage{url}

\title{An overview of nonparametric tests of extreme-value dependence and of some related statistical procedures} 

\author{
   Axel B\"ucher \\
   \small{Fakult\"at f\"ur Mathematik} \\
   \small{Ruhr-Universit\"at Bochum} \\
   \small{44780 Bochum, Germany} \\
   \small{\texttt{axel.buecher@ruhr-uni-bochum.de}}\\
   \and
   Ivan Kojadinovic \\
   \small{Laboratoire de math\'ematiques et applications, UMR CNRS 5142} \\
   \small{Universit\'e de Pau et des Pays de l'Adour} \\
   \small{B.P. 1155, 64013 Pau Cedex, France} \\
   \small{\texttt{ivan.kojadinovic@univ-pau.fr}} 
 }

\begin{document}
\maketitle

\begin{abstract}
An overview of existing nonparametric tests of extreme-value dependence is presented. Given an i.i.d.\ sample of random vectors from a continuous distribution, such tests aim at assessing whether the underlying unknown copula is of the {\em extreme-value} \index{extreme-value copula} type or not. The existing approaches available in the literature are summarized according to how departure from extreme-value dependence is assessed. Related statistical procedures useful when modeling data with this type of dependence are briefly described next. Two illustrations on real data sets are then carried out using some of the statistical procedures under consideration implemented in the \textsf{R} package {\tt copula}. Finally, the related problem of testing the {\em maximum domain of attraction} condition is discussed.
\end{abstract}

\section{Introduction}

By definition, the class of extreme-value copulas consists of all possible limit copulas of affinely normalized, componentwise maxima of a multivariate i.i.d.\ sample, or, more generally, of a multivariate stationary time series. As a consequence, extreme-value copulas can be seen as appropriately capturing the dependence between extreme or rare events. The famous extremal types theorem of multivariate extreme value theory leads to a rather simple characterization of extreme-value copulas: %it states that, under broad conditions, 
the class of extreme-value copulas merely coincides with the class of max-stable copulas (see Section~\ref{bk:sec:found} below for a precise definition). Other characterizations are possible, most of which are based on a parametrization by a lower-dimensional function or measure \citep[see e.g.][for an overview]{bk:GudSeg10}. Serial dependence of the underlying time series is explicitly allowed provided certain mixing conditions hold \citep{bk:Hsi89, bk:Hus90}.

The theory underlying extreme-value copulas motivates their use in combination with the famous {\em block maxima}\index{multivariate block maxima} method popularized in the univariate case in the monograph of \cite{bk:Gum58}: from a given time series, calculate (componentwise) monthly or annual or, more generally, block maxima, and consider the class of extreme-value copulas (or parametric subclasses thereof) as an appropriate model for the multivariate sample of block maxima. If the block size is sufficiently large, it is unlikely that the respective maxima within a block occur at the beginning or the end of the block, whence, even under weak serial dependence of the underlying time series, block maxima could be considered as approximately independent. In statistical practice, independence has usually been postulated hitherto. Applications of the block maxima method can also be found in contexts in which the underlying time series is not necessarily stationary, as is the case when seasonalities are present (for instance in some hydrological problems).

The use of extreme-value copulas is not restricted to the framework of multivariate extreme-value theory. These dependence structures can actually be a convenient choice to model any data sets with positive association. Moreover, many parametric submodels are available in the literature \cite[see e.g.][for an overview]{bk:GudSeg10}.

Extreme-value copulas have been successfully applied in empirical finance and insurance \citep[see e.g.][]{bk:LonSol01, bk:CebDenLam03, bk:McNFreEmb05}, and environmental sciences \citep[see e.g.][]{bk:Taw88,bk:SalDeMKotRos07}. They also arise in spatial statistics in connection with max-stable processes in which they determine the underlying spatial dependence \citep[see e.g.][]{bk:DavPadRib12,bk:Rib13,bk:RibSed13}.

From a statistical point of view, it is important to test the hypothesis that the copula of a given sample is an extreme-value copula. When applied within the context of the block maxima method, a rejection of this hypothesis would indicate that the size of the blocks is too small and should be enlarged, or that the (broad) conditions of the extremal types theorem are not satisfied. When applied outside of the extremal types theoretical framework, tests of extreme-value dependence merely indicate whether the class of max-stable copulas is a plausible choice for modeling the cross-sectional dependence in the data at hand. If there is no evidence against this class, additional statistical procedures tailored for extreme-value copulas can be used to carry out the data analysis.

This chapter is organized as follows. A brief overview of the theory underlying extreme-value copulas is given in the second section. The third section provides a summary of the procedures available in the literature for testing whether the copula of a random sample from a continuous distribution can be considered of the extreme-value type or not. Rather detailed analyses of bivariate financial data and bivariate insurance data are presented next. They are accompanied by code for the \textsf{R} statistical system \citep{bk:Rsystem} from the {\tt copula} package \citep{bk:copula}. Finally, in the last section, the related issue of testing the maximum domain of attraction condition is discussed.

\section{Mathematical foundations} \label{bk:sec:found}

Consider a $d$-dimensional random vector $\mathbf{X} = (X_1,\dots,X_d)$, $d\ge 2$, whose marginal cumulative distributions functions (c.d.f.s) $F_1,\dots,F_d$ are assumed to be continuous. Then, by \cite{bk:Skl59}'s representation theorem, the c.d.f.\ $F$ of $\mathbf{X}$ can be written in a unique way as
$$
F(\mathbf{x}) = C \{ F_1(x_1),\dots,F_d(x_d) \}, \qquad \mathbf{x}=(x_1, \dots, x_d) \in \mathbb{R}^d,
$$
where the function $C:[0,1]^d \to [0,1]$ is a copula, i.e., the restriction of a multivariate c.d.f.\ with standard uniform margins to the unit hypercube. The above display is usually interpreted in the way that the copula $C$ completely characterizes the stochastic dependence among the components of $\mathbf X$.

% Suppose that $\mathbf X_1, \dots, \mathbf X_n$ are independent copies of $\mathbf X$ and consider the vector of componentwise maxima $\mathbf M_n = (M_{n1}, \dots, M_{nd})$, where $M_{nj}= \max(X_{1j}, \dots, X_{nj})$. Then, simple calculations reveal that the c.d.f.\ of $\mathbf M_n$ is $F^n$, and therefore that the corresponding univariate marginal c.d.f.s are $F_1^n, \dots, F_d^n$. After additional steps, the latter implies that the copula of $\mathbf M_n$ evaluated at $\mathbf{u} \in [0,1]^d$ is $\{ C(u_1^{1/n}, \dots, u_1^{1/n}) \}^n$. Indeed,
% \begin{align*}
% C^*(\mathbf{u}) &= \left( F \left[ \{ F_1^n \}^{-1}(u_1), \dots, \{ F_d^n \}^{-1}(u_1) \right] \right)^n \\
% &= F\left[ \left\{ F_1^{-1}(u_1^{1/n}), \dots, F_1^{-1}(u_1^{1/n}) \right\} \right]^n\\
% &= \left\{ C(u_1^{1/n}, \dots, u_1^{1/n}) \right\}^n, \qquad \mathbf{u} \in [0,1]^d.
% \end{align*}

A $d$-dimensional copula $C$ is an {\em extreme-value} copula \index{extreme-value copula} if and only if there exists a copula $C^*$ such that, for any $\mathbf{u} \in [0,1]^d$,
\begin{align}
\label{bk:eq:maxdom}
\lim_{n \to \infty} \{ C^*(u_1^{1/n},\dots,u_d^{1/n}) \}^n = C(\mathbf{u}).
\end{align}
The copula $C^*$ is then said to be in the {\em maximum domain of attraction} of $C$, which shall be denoted as $C^* \in D(C)$ in what follows.

Some algebra reveals that $\{ C^*(u_1^{\scriptscriptstyle 1/n},\dots,u_d^{\scriptscriptstyle  1/n}) \}^n$ is the copula, evaluated at $\mathbf{u} \in [0,1]^d$, of the vector of componentwise maxima computed from an i.i.d.\ sample $\mathbf Y_1, \dots, \mathbf Y_n$ with continuous marginal c.d.f.s and copula $C^*$. The latter fact motivates the terminology \textit{extreme-value copula}. 
 It is additionally very useful to note that $C$ is an extreme-value copula if and only if it is {\em max-stable}, that is, if and only if, for any $\mathbf{u} \in [0,1]^d$ and $r  \in  \mathbb N$, $r > 0$,
\begin{equation}
\label{bk:eq:maxstability}
\{ C(u_1^{1/r},\dots,u_d^{1/r}) \}^r = C(\mathbf{u}).
\end{equation}
The sufficiency follows by using, in combination with~\eqref{bk:eq:maxdom}, the fact that, for any $\mathbf{u} \in [0,1]^d$ and $r \in\mathbb N$, $r > 0$,
\[
\Big[C^*\big\{ (u_1^{1/r})^{1/n}, \dots, (u_d^{1/r})^{1/n} \big\} \Big]^{1/n}
=
\Big[ \big \{ C^*(u_1^{1/(nr)}, \dots, u_d^{1/(nr)}) \big\}^{1/(nr)} \Big]^r.
\]
The necessity is an immediate consequence of the fact $C \in D(C)$ for any max-stable copula $C$. Interestingly enough, it can be shown that a max-stable copula actually satisfies~\eqref{bk:eq:maxstability} for any real $r>0$ \citep[see e.g.][Lemma 5.4.1]{bk:Gal78}. \index{max-stable copula}

An alternative, more complex characterization, essentially due to \cite{bk:Pic81}, is as follows: a copula $C$ is of the extreme-value type if and only if there exists a function $A$ such that, for any $\mathbf{u} \in (0,1]^d \setminus \{(1,\dots,1)\}$,
\begin{equation}
\label{bk:eq:Pickands_charact}
C(\mathbf{u})
=
\exp \left\{ \left( \sum_{j=1}^d \log u_j \right) A \left(\frac{\log u_2}{\sum_{j=1}^d \log u_j}, \dots, \frac{\log u_{d}}{\sum_{j=1}^d \log u_j} \right) \right\},
\end{equation}
where $A:\Delta_{d-1} \to [1/d,1]$ is the {\em Pickands dependence function} \index{Pickands dependence function} and $\Delta_{d-1} = \{(w_1,\dots,w_{d-1}) \in [0,1]^{d-1} : w_1 + \dots + w_{d-1} \le 1 \}$ is the unit simplex \citep[see e.g.][for more details]{bk:GudSeg12}. If relation~\eqref{bk:eq:Pickands_charact} is met, then $A$ is necessarily convex and satisfies the boundary condition $\max\{1- \sum_{j=1}^{d-1} w_j, w_1, \dots, w_{d-1}\} \le A(\mathbf w) \le 1$ for all $\mathbf w = (w_1,\dots,w_{d-1}) \in \Delta_{d-1}$. The latter two conditions are, however, not sufficient to characterize the class of Pickands dependence functions unless $d=2$ \citep[see e.g.][for a counterexample]{bk:BeiGoeSegTeu04}.

Several other characterizations of extreme-value copulas are possible, for instance using the {\em spectral measure of $C$} \citep[see e.g.][for details]{bk:GudSeg12} or the {\em stable tail dependence function} \citep{bk:res13, bk:ChaFouGenNes14}. 

\section{Existing tests of extreme-value dependence}
\label{bk:sec:tests}

Let $\mathcal{EV}$ denote the class of extreme-value copulas. Given a random sample $\mathbf{X}_1,\dots,\mathbf{X}_n$ from a c.d.f.\ $C\{F_1(x_1),\dots,F_d(x_d)\}$ with $F_1,\dots,F_d$ continuous and $C,F_1,\dots,F_d$ unknown, tests of extreme-value dependence aim at testing
\begin{equation}
\label{bk:eq:H0}
H_0 : C \in \mathcal{EV} \qquad \mbox{against} \qquad H_1 : C \not \in \mathcal{EV}.
\end{equation}
The existing tests for $H_0$ available in the literature are all rank-based and therefore margin-free. They can be classified into three groups according to how departure from extreme-value dependence is assessed. 

\subsection{Approaches based on Kendall's distribution} \label{bk:subsec:kendall}

The first class of approaches, which is also the oldest, finds its origin in the seminal work of \cite{bk:GhoKhoRiv98} and is restricted to the case $d=2$. Given a bivariate random vector $\mathbf{X} = (X_1,X_2)$ with c.d.f.\ $F$, continuous marginal c.d.f.s $F_1$ and $F_2$ and copula $C$, the tests in this class are based on the random variable
$$
W = F(X_1,X_2) = C \{ F_1(X_1), F_2(X_2) \}.
$$
The c.d.f.\ of $W$ is frequently referred to as {\em Kendall's distribution} and will be denoted by $K$ subsequently. When $C \in \mathcal{EV}$, \cite{bk:GhoKhoRiv98} showed that
\begin{equation}
\label{bk:eq:Kendall_dist}
K(w) = \Pr(W \leq w) = w - (1 - \tau) w \log w, \qquad w \in (0,1],
\end{equation}
where $\tau$ denotes {\em Kendall's tau}. Whether $C$ is of the extreme-value type or not, it is known since \cite{bk:SchWol81} that
$$
\tau = 4 \int_{[0,1]^2} C(u_1,u_2) \mathrm{d} C(u_1,u_2) - 1 = 4 \mathrm{E}(W) - 1.
$$
When $C \in \mathcal{EV}$, \cite{bk:GhoKhoRiv98} also obtained from~\eqref{bk:eq:Kendall_dist} that, for $k\in \mathbb N$,
$
\mu_k := E(W^k) = (k \tau + 1)/(k+1)^2, 
$
which for instance implies that
\begin{equation}
\label{bk:eq:test1}
-1 + 8 \mu_1 - 9 \mu_2 = 0.
\end{equation}
In order to test $H_0$ from a bivariate random sample $\mathbf{X}_1,\dots,\mathbf{X}_n$ with c.d.f.\ $C\{F_1(x_1),F_2(x_2)\}$ where $C,F_1,F_2$ are unknown, \cite{bk:GhoKhoRiv98} suggested to assess whether a sample version of the left-hand side of~\eqref{bk:eq:test1} is significantly different from zero or not. Specifically, they considered the statistic
\begin{equation}
\label{bk:eq:S2n}
S_{2n} = - 1 + \frac{8}{n(n-1)} \sum_{i \neq j} I_{ij} - \frac{9}{n(n-1)(n-2)} \sum_{i \neq j \neq k} I_{ij} I_{kj},
\end{equation}
where $I_{ij} = \mathbf{1}(X_{i1} \leq X_{j1},X_{i2} \leq X_{j2})$. As shown by \cite{bk:GhoKhoRiv98}, $S_{2n}$ is a centered $U$-statistic which, under the null hypothesis, converges weakly to a normal random variable. To carry out the test, \cite{bk:GhoKhoRiv98} proposed to estimate the variance of $S_{2n}$ using a jackknife estimator. The test based on $S_{2n}$ was revisited by \cite{bk:BenGenNes09} who proposed two alternative strategies to compute approximate p-values for $S_{2n}$. The three versions of the test are implemented in the function \texttt{evTestK} of the \textsf{R} package {\tt copula}.

The above approach was recently furthered by \cite{bk:DuNes13} who used the first three moments of Kendall's distribution and the theoretical relationship
\begin{equation}
\label{bk:eq:test2}
-1+4\mu_1+9\mu_2-16\mu_3 = 0
\end{equation}
under the null instead of~\eqref{bk:eq:test1}. The corresponding test statistic will subsequently be denoted by $S_{3n}$.
An additional contribution of the latter authors was to find a counterexample to \cite{bk:GhoKhoRiv98}'s conjecture that $K$ has the form in~\eqref{bk:eq:Kendall_dist} if and only if $C \in \mathcal{EV}$. The latter implies that tests in this class are not consistent. Despite that fact, the Monte Carlo experiments reported in \cite{bk:KojYan10c} and in \cite{bk:DuNes13} suggest that tests based on $S_{2n}$ and its extension studied in \cite{bk:DuNes13} are among the most powerful procedures for testing bivariate extreme-value dependence.

Notice finally that additional extensions of the approach of \cite{bk:GhoKhoRiv98} were studied in \cite{bk:Que12} along with tests based on Cram\'er--von Mises-like statistics derived from the empirical process $\sqrt{n} (K_n - K_{\tau_n})$, where $K_n$ is the empirical c.d.f.\ of $\hat W_1,\dots,\hat W_n$ with $\hat W_i = F_n(X_{i1},X_{i2})$ and $F_n$ the empirical c.d.f.\ of $\mathbf{X}_1,\dots,\mathbf{X}_n$, and $K_{\tau_n}$ is defined as in~\eqref{bk:eq:Kendall_dist} with $\tau$ replaced by its classical estimator denoted $\tau_n$.

\subsection{Approaches based on max-stability} \label{bk:subsec:max}

The second class of tests proposed in the literature consists of assessing empirically whether~\eqref{bk:eq:maxstability} holds or not. It was investigated in \cite{bk:KojSegYan11} for $d \geq 2$. The key ingredient is a natural nonparametric estimator of the unknown copula $C$ known as the {\em empirical copula} \citep[see e.g.][]{bk:Rus76,bk:Deh79,bk:Deh81}.

Given a sample $\mathbf{X}_1,\dots,\mathbf{X}_n$ from a c.d.f.\ $C\{F_1(x_1),\dots,F_d(x_d)\}$ with $F_1,\dots,F_d$ continuous and $C,F_1,\dots,F_d$ unknown, let $\hat U_{ij} = R_{ij}/(n+1)$ for all $i \in \{1,\dots,n\}$ and $j \in \{1,\dots,d\}$, where $R_{ij}$ is the rank of $X_{ij}$ among $X_{1j},\dots,X_{nj}$, and set $\mathbf{\hat U}_i = (\hat U_{i1},\dots, \hat U_{id})$. It is worth noticing that the scaled ranks $\hat U_{ij}$ can equivalently be rewritten as $\hat U_{ij} = n F_{nj}(X_{ij}) / (n+1)$, where $F_{nj}$ is the empirical c.d.f.\ computed from $X_{1j},\dots,X_{nj}$, the scaling factor $n/(n+1)$ being classically introduced to avoid problems at the boundary of~$[0,1]^d$. The empirical copula of $\mathbf{X}_1,\dots,\mathbf{X}_n$ is then frequently defined as the empirical c.d.f.\ computed from the {\em pseudo-observations} \index{pseudo-observations} $\mathbf{\hat U}_1,\dots,\mathbf{\hat U}_n$, i.e.,
\begin{equation}
\label{bk:eq:empcop}
C_n(\mathbf{u}) = \frac{1}{n} \sum_{i=1}^n \mathbf{1} ( \mathbf{\hat U}_i \leq \mathbf{u} ), \qquad \mathbf{u} \in [0,1]^d.
\end{equation}
The inequalities between vectors in the above definition are to be understood componentwise.

To test~\eqref{bk:eq:maxstability} empirically, \cite{bk:KojSegYan11} considered test statistics constructed from the empirical process
\begin{equation}
\label{bk:test_process}
\mathbb{D}_{r,n}(\mathbf{u}) = \sqrt{n} \left[ \{ C_n(u_1^{1/r},\dots,u_d^{1/r}) \}^r - C_n(\mathbf{u}) \right], \qquad \mathbf{u} \in [0,1]^d,
\end{equation}
for some strictly positive fixed values of $r$. The recommended test statistic is
\begin{equation}
\label{bk:eq:T345n}
T_{3,4,5,n} = T_{3,n} + T_{4,n} + T_{5,n},
\end{equation}
where $T_{r,n} = \int_{[0,1]^d} \{\mathbb{D}_{r,n}(\mathbf{u})\}^2 \mathrm{d} C_n(\mathbf{u})$. Approximate p-values for the latter were computed using a {\em multiplier bootstrap}. The test based on $T_{3,4,5,n}$ is implemented in the function \texttt{evTestC} of the \textsf{R} package {\tt copula}. It is not a consistent test either, because the validity of~\eqref{bk:eq:maxstability} is assessed only for a small number of $r$ values.

\subsection{Approaches based on the estimation of the Pickands dependence function}
\label{bk:subsec:pick}

Recall that $\mathbf{X}_1,\dots,\mathbf{X}_n$ is a random sample from a c.d.f.\ $C\{F_1(x_1),\dots,F_d(x_d)\}$ with $F_1,\dots,F_d$ continuous and $C,F_1,\dots,F_d$ unknown. If $C \in \mathcal{EV}$, it can be expressed as in~\eqref{bk:eq:Pickands_charact}. The third class of tests exploits variations of the following idea: given a nonparametric estimator $A_n$ of $A$ and using the empirical copula $C_n$ defined in~\eqref{bk:eq:empcop}, relationship~\eqref{bk:eq:Pickands_charact} can be tested empirically.

The first test in this class is due to \cite{bk:KojYan10c} who, for $d=2$ only, constructed test statistics from the empirical process
$$
\mathbb{E}_n(u_1,u_2) = \sqrt{n} \left( C_n(u_1,u_2) - \exp \left[ \log(u_1 u_2) A_n \left\{ \frac{\log(u_2)}{\log(u_1 u_2)} \right\} \right] \right),
$$
for $(u_1,u_2) \in (0,1]^2 \setminus \{(1,1)\}$. The recommended statistic is
\begin{equation}
\label{bk:eq:TnA}
T_n^A = \int_{[0,1]^2} \mathbb{E}_n(u_1,u_2)^2 \mathrm{d} C_n(u_1,u_2),
\end{equation}
when $A_n$ is the rank-based version of the Cap\'era\`a--Foug\`eres--Genest (CFG) estimator of $A$ studied in \cite{bk:GenSeg09}. The resulting test relies on a {\em multiplier bootstrap} and is implemented in the function \texttt{evTestA} of the \textsf{R} package {\tt copula}. A multivariate version of this test was studied in \cite{bk:Gud12} using the multivariate extension of the rank-based CFG estimator of~$A$ investigated in \cite{bk:GudSeg12}.

An alternative class of nonparametric multivariate rank-based estimators of $A$ was proposed in \cite{bk:BucDetVol11} and \cite{bk:BerBucDet13}. These are based on the minimization of a weighted $L^2$-distance between the logarithms of the empirical and the unknown extreme-value copula. To derive multivariate tests of extreme-value dependence, the latter authors reused the aforementioned $L^2$-distance to measure the difference between the empirical copula in~\eqref{bk:eq:empcop} and a plug-in nonparametric estimator of $C$ under extreme-value dependence based on~\eqref{bk:eq:Pickands_charact}. The corresponding test statistic is subsequently denoted by $T_{L^2,n}$.

We end this subsection by briefly summarizing a recent graphical approach due to \cite{bk:CorGenNes14}. Their idea, hitherto restricted to the bivariate case, is as follows: given a copula $C$, consider the transformation $T_C:(0,1)^2 \to (0,1) \times (0,\infty]$, defined by
\[
T_C(u_1,u_2) =  \left( \frac{\log (u_2)}{\log (u_1u_2)}, \frac{\log \{ C(u_1,u_2) \} }{ \log(u_1u_2) } \right), \qquad (u_1,u_2) \in (0,1)^2.
\]
If $C \in \mathcal{EV}$, representation~\eqref{bk:eq:Pickands_charact} holds and we have $\log \{ C(u_1,u_2) \} = \log(u_1u_2) A \{ \log(u_2)/\log(u_1u_2) \}$ for all $(u_1,u_2) \in (0,1)^2$, whence $\mathcal S_C = \{ T_C(u,v): (u,v) \in (0,1)^2\}$ coincides with the graph of $A$, i.e., with the set $\{(t,A(t)) : t \in (0,1) \}$. More generally, some thought reveals that $H_0$ is valid if and only if $\mathcal  S_C$ is a convex curve. The latter observation suggests to test $H_0$ in~\eqref{bk:eq:H0} by estimating the set $\mathcal S_C$ and visually assessing the departure of that estimated set from a convex curve. The estimator defined in \cite{bk:CorGenNes14}, called the {\em A-plot}, is given by
\[
\hat {\mathcal S}_n =
\left\{
(\hat T_i,  \hat Z_i) : \hat T_i = \frac{\log(\hat U_{i2}) }{ \log( \hat U_{i1} \hat U_{i2})}, \hat Z_i = \frac{\log \{ C_n(\hat U_{i1}, \hat U_{i2}) \} }{ \log(\hat U_{i1} \hat U_{i1})}, i \in \{1, \dots, n\}
\right\}.
\]
Examples of A-plots when $C \in \mathcal{EV}$ and when $C \not \in \mathcal{EV}$ can be found in Figure~1 of \cite{bk:CorGenNes14}. When $C$ is of the extreme-value type, the previous authors proposed a B-spline smoothing estimator for the Pickands dependence function $A$ based on $\hat {\mathcal S}_n$. The latter is subsequently denoted by $A_n$ for simplicity (even though the estimator depends on several smoothing parameters). Additionally to a pure graphical check, the authors propose
\begin{equation}
\label{bk:eq:resid}
T_n = \frac{1}{n} \sum_{i=1}^n  \{ \hat Z_i - A_n (\hat T_i) \}^2,
\end{equation}
a residual sum of squares, as a formal test statistic for $H_0$. The hypothesis is rejected for unlikely large values of $T_n$. Specifically, an approximate p-value for $T_n$ is computed by means of a {\em parametric bootstrap} procedure based on simulating from a copula with Pickands dependence function $A_n$.

\subsection{Finite-sample performance of some of the tests}

The finite-sample performance of the tests reviewed in the preceding sections was investigated by various authors. Table~\ref{bk:tab:evc} below, taken from \cite{bk:CorGenNes14}, gathers those results from \cite{bk:CorGenNes14}, \cite{bk:KojYan10c}, \cite{bk:DuNes13}, and \cite{bk:BucDetVol11} that were obtained under the same experimental settings (notice that the Gumbel--Hougaard copula is the only extreme-value copula among those considered in the table). As noted by \cite{bk:CorGenNes14}, no test is uniformly better than the others: each test, except the one based on $T_{L^2,n}$ from \cite{bk:BucDetVol11}, is favored for at least one of the considered scenarios under $H_1$. For high levels of dependence (as measured by Kendall's tau), the tests based on $S_{2n}$ and $S_{3n}$ described in Section~\ref{bk:subsec:kendall} seem to yield the most accurate approximation of the nominal level (here 5\%). The tests whose approximate p-values are computed by means of a multiplier bootstrap, i.e., the tests based on $T_{3,4,5,n}$ defined in~\eqref{bk:eq:T345n} and on $T_n^A$ and $T_{L^2,n}$ introduced in Section~\ref{bk:subsec:pick}, are quite conservative for such scenarios. From a computational perspective, the test based on $S_{2n}$ seems to be the fastest, while the one based on $T_n^A$ defined in~\eqref{bk:eq:TnA} is the most computationally intensive. Additional comparison of the tests based on $S_{2n}$, $T_{3,4,5,n}$ and $T_n^A$ (resp.\ $S_{2n}$ and $S_{3n}$) can be found in \citet[Tables 1--3]{bk:KojYan10c} \citep[resp.][Table 5]{bk:DuNes13}.

\begin{table}[t!]
    \begin{center}
    \footnotesize{
        \begin{tabular}{l l r  r  r  r  r  r}
        $\tau$ & $C$ & $T_n$ & $S_{2n}$ & $S_{3n}$ & $T_n^A$ & $T_{3,4,5,n}$ & $T_{L^2,n}$ \\  \hline 
        0.25 & Gumbel--Hougaard & 4.7 & 5.4 & 5.3 & 3.8 & 5.0 & 4.5 \\
        & Clayton & 97.7 & 98.0 & 96.6 & {\bf 98.4} & 94.6 & 87.4 \\
        & Frank & 18.7 & 38.4 & 57.0 & 58.3 & {\bf 66.1} & 29.1 \\
        & Gaussian & 25.5 & 37.3 & {\bf 40.3} & 36.5 & 38.7 & 16.8 \\
        & Student $t$ with 4 d.f.\ & {\bf 37.7} & 26.2 & 19.6 & 23.9 & 26.6 & 10.5 \\ \hline
        0.5 & Gumbel--Hougaard & 5.4 & 5.1 & 5.0 & 3.9 & 4.0 & 2.9 \\
        & Clayton & 100.0 & 100.0 & 100.0 & 100.0 & 100.0 & 100.0 \\
        & Frank & 87.8 & 59.4 & 84.4 & 95.7 & {\bf 96.5} & 73.0 \\
        & Gaussian & 59.4 & {\bf 62.6} & 61.7 & 61.8 & 51.0 & 23.7 \\
        & Student $t$ with 4 d.f.\ & {\bf 58.6} & 56.0 & 45.3 & 50.1 & 52.7 & 15.8 \\ \hline
        0.75 & Gumbel--Hougaard & 6.2 & 4.9 & 5.3 & 3.2 & 2.3 & 2.5 \\
         & Clayton & 100.0 & 100.0 & 100.0 & 100.0 & 100.0 & 100.0 \\
         & Frank & 98.3  & 58.5 & 92.9 & {\bf 99.9} & 99.0 & 78.3 \\
         & Gaussian & 56.5 & {\bf 75.2} & 71.1 & 66.5 & 46.7 & 8.4 \\
         & Student $t$ with 4 d.f.\ & 45.8 & 67.8 & 55.8 & 50.6 & {\bf 69.2} & 4.6 \\
        \end{tabular}
    }
    \caption{\small Rejection rates of $H_0$ estimated from random samples of size $n=200$ generated from a c.d.f.\ with copula $C$ whose Kendall's tau is~$\tau$. All the tests were carried out at the 5\% significance level. The table is taken from \cite{bk:CorGenNes14}.
    }\label{bk:tab:evc}
    \end{center}
        \vspace{-.7cm}
\end{table}

\cite{bk:KojSegYan11} and \cite{bk:BerBucDet13} also present 
simula\-tion results for $d > 2$, which are in favor of the test based on $T_{3,4,5,n}$ defined in~\eqref{bk:eq:T345n}. Preliminary results obtained in \cite{bk:Gud12} indicate that the multivariate extension of the test based on $T_n^A$ defined in~\eqref{bk:eq:TnA} is likely to outperform the test based on $T_{3,4,5,n}$ for several scenarios under $H_1$.

\section{Some related statistical inference procedures}
\label{bk:sec:related}

Once it has been decided to use an extreme-value copula to model dependence in a set of multivariate continuous i.i.d.\ observations, a typical next step is to choose a parametric family $\mathcal{C}$ in $\mathcal{EV}$ and estimate its unknown parameter(s) from the data. As many parametric families of extreme-value copulas are available \citep[see e.g.][]{bk:GudSeg10,bk:RibSed13}, it is of strong practical interest to be able to test whether a given family $\mathcal{C}$ is a plausible model or not for the data at hand. In other words, tests for
%$$
%H_0 : C \in \mathcal{C} \qquad \mbox{against} \qquad H_1 : C \not \in \mathcal{C}
%$$
$
H_0 : C \in \mathcal{C}
$  
against  
$
H_1 : C \not \in \mathcal{C}
$
would be needed. Such goodness-of-fit procedures were investigated in the bivariate case by \cite{bk:GenKojNesYan11} who considered Cram\'er--von Mises test statistics based on the difference between a nonparametric and a parametric estimator of the Pickands dependence function. The Monte Carlo experiments reported in the latter work highlighted the fact that, unless the amount of data is very large, there is hardly any practical difference among the existing bivariate symmetric parametric families of extreme-value copulas, and that an issue of more importance from a modeling perspective is whether a symmetric or asymmetric family should be used. For that purpose, the specific test of symmetry for bivariate extreme-value copulas investigated in \cite{bk:KojYan12} can be used as a complement to the goodness-of-fit test studied in \cite{bk:GenKojNesYan11}. Both tests are available in the {\tt copula} \textsf{R} package.

When $d > 2$ but $d$ remains reasonably small (say $d \leq 10$), generic goodness-of-fit tests (that is, developed for any parametric copula family, not necessarily of the extreme-value type) could be used \citep[see e.g.][and the references therein]{bk:GenRemBea09,bk:KojYan11}. In a higher dimensional context, one possibility consists of using the specific approach for extreme-value copulas proposed by \cite{bk:Smi90} in his seminal work on max-stable processes. It consists of comparing nonparametric and parametric estimators of the underlying {\em extremal coefficients} (which are functionals of the Pickands dependence function). The latter approach was recently revisited in \cite{bk:KojShaYan14}.

\section{Illustrations and \textsf{R} code from the {\tt copula} package}
\label{bk:sec:illus}

We provide two illustrations below. The first one concerns bivariate financial logreturns and exemplifies the key theoretical connection between multivariate block maxima and extreme-value copulas briefly mentioned in the introduction and Section~\ref{bk:sec:found}. The second illustration consists of a detailed analysis of the well-known LOSS/ALAE insurance data with particular emphasis on the effect and handling of ties.

\subsection{Bivariate financial logreturns}

As a first illustration, we considered daily logreturns computed from the closing values of the Dow Jones and the S\&P 500 stock indexes for the period 1990-2004. The closing values are available in the {\tt QRM} \textsf{R} package \citep{bk:QRM} and can be loaded by entering the following commands into the \textsf{R} terminal:
{\small 
\begin{verbatim}
> library(QRM)
> data(dji)
> data(sp500)
\end{verbatim}
} \noindent
Daily logreturns for the period under consideration were computed using the {\tt timeSeries} \textsf{R} package \citep{bk:timeSeries}:
{\small
\begin{verbatim}
> d <- na.omit(cbind(dji,sp500))
> rd <- returns(d)
\end{verbatim}
}\noindent
The statistical procedures mentioned in the previous sections should not however be directly applied on the resulting bivariate daily logreturns as the latter are strongly serially dependent. To obtain observations that might exhibit extreme-value dependence and could be considered approximately i.i.d., we first formed the bivariate series of componentwise monthly maxima. The last step was performed using functions from the {\tt timeSeries} and {\tt timeDate} \textsf{R} packages \citep{bk:timeDate}:
{\small
\begin{verbatim}
> by <- timeSequence(from=start(rd),  to=end(rd), by="month")
> mrd <- aggregate(rd, by, max)
\end{verbatim}}
\noindent The resulting component series do not contain ties which is compatible with the implicit assumption of continuous margins:
{\small
\begin{verbatim}
> x <- series(mrd)
> nrow(x)
[1] 171
> apply(x, 2, function(x) length(unique(x)))
  DJI SP500 
  171   171 
\end{verbatim}
} \noindent
After loading the {\tt copula} package with the command {\tt library(copula)} and setting the random seed by typing {\tt set.seed(123)}, the test of extreme-value dependence based on $S_{2n}$ (resp.\ $T_{3,4,5,n}$, $T_n^A$) defined in~\eqref{bk:eq:S2n} (resp.\,~\eqref{bk:eq:T345n}, ~\eqref{bk:eq:TnA}) was applied using the command {\tt evTestK(x)} (resp. {\tt evTestC(x)}, {\tt evTestA(x, derivatives="Cn")}) and returned an approximate p-value of 0.5737 (resp.\ 0.4191, 0.2423). In other words, none of the tests detected any evidence against extreme-value dependence thereby suggesting that the copula of componentwise block maxima, for blocks of length corresponding to a month, is sufficiently close to an extreme-value copula. Note that, as the tests are rank-based, they could have equivalently been called on the pseudo-observations computed from the monthly block maxima. The random seed was set (to ensure exact reproducibility) because the second and third tests involve random number generation as their p-values are computed using resampling.

For illustration purposes, we next formed monthly logreturns as follows:
{\small
\begin{verbatim}
> srd <- aggregate(rd, by, sum)
> x <- series(srd)
\end{verbatim}
} \noindent
Proceeding as previously, it can be verified that the resulting component series do not contain ties which is compatible with the implicit assumption of continuous margins. Monthly log\-returns being merely sums of daily log\-returns, the underlying unknown bivariate distribution should be far from exhibiting extreme-value dependence. The tests of extreme-value dependence based on $S_{2n}$, $T_{3,4,5,n}$ and $T_n^A$ returned approximate p-values of 0.0003, 0.02 and 0.0005, respectively, confirming that there is strong evidence in the data against extreme-value dependence. 

\subsection{LOSS/ALAE insurance data}

The well-known LOSS/ALAE insurance data are very frequently used for illustration purposes in copula modeling \citep[see e.g.][]{bk:FreVal98,bk:BenGenNes09,bk:KojYan10}. The two variables of interests are LOSS, an indemnity payment, and ALAE, the corresponding allocated loss adjustment expense. They were observed for 1500 claims of an insurance company. Following \cite{bk:BenGenNes09}, the following study is restricted to the 1466 uncensored claims.

The data are available in the {\tt copula} package, and can be loaded by typing {\tt library(copula)} followed by {\tt data(loss)}. The uncensored claims described in terms of LOSS and ALAE were obtained as follows:
{\small
\begin{verbatim}
> myLoss <- subset(loss, censored==0, select=c("loss", "alae"))
\end{verbatim}
}\noindent
These data, consisting of 1466 bivariate observations, contain a non-negligible amount of ties, the variable LOSS being particularly affected:
{\small
\begin{verbatim}
> sapply(myLoss, function(x) length(unique(x)))
loss alae
 541 1401
\end{verbatim}
} 

The presence of ties is incompatible with the implicit assumption of continuous margins. Indeed, combined with the assumption that the data are i.i.d.\ observations, continuity of the margins implies that ties should no occur. Yet, ties are present here as in many other real data sets. The latter could be due either to the fact that the observed phenomena are truly discontinuous, or to precision/rounding issues. As far as the LOSS/ALAE data are concerned, the latter explanation applies.

Among the tests briefly described in Section~\ref{bk:sec:tests}, only that of \cite{bk:CorGenNes14} explicitly considers the case of discontinuous margins (see Section~6 in that reference). The remaining tests were all implemented under the assumption of continuous margins. For the test based on $S_{2n}$ defined in~\eqref{bk:eq:S2n}, \cite{bk:GenNesRup11} provide a heuristic explanation of the fact that, for discontinuous margins, $S_{2n}$ is not necessarily centered anymore under the null.

Given the situation, there are roughly four possible courses of action: (i)~stop the analysis, (ii)~delete tied observations, (iii)~use average ranks for ties or (iv)~break ties at random, sometimes referred to as {\em jittering} (which amounts to adding a small, continuous white noise term to all observations). Arguments for not considering solution~(ii) are given in~\citet[Section~2]{bk:GenNesRup11}. To empirically study solutions~(iii) and~(iv), the latter authors carried out an experiment consisting of applying  the test based on $S_{2n}$ defined in~\eqref{bk:eq:S2n} on binned observations from a bivariate Gumbel--Hougaard copula. More specifically, tied observations were obtained by dividing the unit square uniformly into bins of dimension 0.1 by 0.1 (resp.\ 0.2 by 0.2) resulting in at most 100 (resp.\ 25) different bivariate observations whatever the sample size. In such a setting, \cite{bk:GenNesRup11} observed that both solutions~(iii) and~(iv) led to strongly inflated empirical levels for the test based on $S_{2n}$.

The situation in terms of ties in the LOSS/ALAE data is however far from being as extreme as in the experiment of \cite{bk:GenNesRup11}. In addition, ties mostly affect the LOSS variable. This prompted us first to consider solution~(iv) as implemented in \cite{bk:KojYan10}. 

\paragraph{Random ranks for ties} The idea consists of carrying out the analysis for many different randomizations (with the hope that this will result in many different configurations for the parts of the data affected by ties) and then looking at the empirical distributions (and not the averages) of the results (here the p-values of various tests).  

For illustration purposes, we first detail the analysis for one randomization:
{\small
\begin{verbatim}
> set.seed(123)
> pseudoLoss <- sapply(myLoss, rank, ties.method="random") /
                (nrow(myLoss) + 1)
\end{verbatim}
} \noindent
As a next step, the tests of extreme-value dependence based on $S_{2n}$, $T_{3,4,5,n}$ and $T_n^A$ defined in~\eqref{bk:eq:S2n},~\eqref{bk:eq:T345n} and~\eqref{bk:eq:TnA}, respectively, were applied by successively typing \texttt{evTestK(pseudoLoss)}, \texttt{evTestC(pseudoLoss)} and \texttt{evTestA(pseudoLoss, derivatives="Cn")}, resulting in approximate p-values of 0.8845, 0.468 and 0.4231, respectively. 
%{\small
%\begin{verbatim}
%> evTestK(pseudoLoss)
%
%	Test of bivariate extreme-value dependence based on Kendall's process
%	with argument 'method' set to “fsample”
%
%data:  pseudoLoss
%statistic = -0.1453, p-value = 0.8845
%
%> evTestC(pseudoLoss)
%
%	Max-stability based test of extreme-value dependence for multivariate
%	copulas
%
%data:  pseudoLoss
%statistic = 0.1469, p-value = 0.468
%
%> evTestA(pseudoLoss, derivatives="Cn")
%
%	Test of bivariate extreme-value dependence based on the CFG estimator
%	with argument 'derivatives' set to 'Cn'
%
%data:  pseudoLoss
%statistic = 0.0186, p-value = 0.4231
%\end{verbatim}}
Hence, none of the three tests detected any evidence against extreme-value dependence. The following step consisted of fitting a parametric family of bivariate extreme-value copulas to the data. As discussed in Section~\ref{bk:sec:related}, given the very strong similarities among the existing families of bivariate symmetric extreme-value copulas, the only issue of practical importance is to assess whether a symmetric or asymmetric family should be used. To do so, we applied the test developed in \cite{bk:KojYan12} by calling \texttt{exchEVTest(pseudoLoss)}, with a resulting p-value of 0.1653.
%{\small
%\begin{verbatim}
%> exchEVTest(pseudoLoss)
%
%	Test of exchangeability for bivariate extreme-value copulas with
%	argument 'estimator' set to 'CFG', argument 'derivatives' set to 'Cn'
%	and argument 'm' set to 100
%
%data:  pseudoLoss
%statistic = 0.0751, p-value = 0.1653
%\end{verbatim}}
The previous result suggested to focus on an exchangeable family such as the Gumbel--Hougaard. We then ran the goodness-of-fit test proposed in \cite{bk:GenKojNesYan11}
by calling 
{\small
\begin{verbatim}
> gofEVCopula(gumbelCopula(), pseudoLoss, method="itau", verbose=FALSE)
\end{verbatim}}
\noindent
The resulting p-value of 0.2592 suggested to fit the Gumbel--Hougaard family:
%{\small
%\begin{verbatim}
%> gofEVCopula(gumbelCopula(), pseudoLoss, method="itau", verbose=FALSE)
%
%	Parametric bootstrap based GOF test for EV copulas with argument
%	'method' set to 'itau' and argument 'estimator' set to 'CFG'
%
%data:  pseudoLoss
%statistic = 0.0377, parameter = 1.44, p-value = 0.2592
%\end{verbatim}}
%This last p-value suggested to fit the Gumbel--Hougaard family:
{\small
\begin{verbatim}
> fitCopula(gumbelCopula(), pseudoLoss, method="itau")
fitCopula() estimation based on 'inversion of Kendall's tau'
and a sample of size 1466.
      Estimate Std. Error z value Pr(>|z|)
param  1.44040    0.03327   43.29   <2e-16 ***
---
Signif. codes:  0 ‘***’ 0.001 ‘**’ 0.01 ‘*’ 0.05 ‘.’ 0.1 ‘ ’ 1
\end{verbatim}}

To assess how different randomizations of the ties affect the results, the above analysis was repeated 100 times using the following code:
{\small
\begin{verbatim}
> randomize <- function()
+ {
+     pseudoLoss <- sapply(myLoss, rank, ties.method="random") /
                    (nrow(myLoss) + 1)
+     evtK <- evTestK(pseudoLoss)$p.value
+     evtC <- evTestC(pseudoLoss)$p.value
+     evtA <- evTestA(pseudoLoss, derivatives="Cn")$p.value
+     exevt <- exchEVTest(pseudoLoss)$p.value
+     gofevGH <- gofEVCopula(gumbelCopula(), pseudoLoss, method="itau",
                             verbose=FALSE)$p.value
+     fitGH <- fitCopula(gumbelCopula(), pseudoLoss, method="itau")
+     c(evtK=evtK, evtC=evtC, evtA=evtA, exevt=exevt, gofevGH=gofevGH,
+       est=fitGH@estimate, se=sqrt(fitGH@var.est))
+ }
> reps <- t(replicate(100, randomize()))
> round(apply(reps, 2, summary), 3)
         evtK  evtC  evtA exevt gofevGH   est    se
Min.    0.868 0.430 0.353 0.092   0.191 1.441 0.033
1st Qu. 0.898 0.462 0.396 0.112   0.223 1.442 0.033
Median  0.914 0.475 0.411 0.120   0.235 1.442 0.033
Mean    0.913 0.474 0.411 0.122   0.236 1.442 0.033
3rd Qu. 0.928 0.489 0.425 0.129   0.248 1.443 0.033
Max.    0.955 0.525 0.464 0.162   0.292 1.444 0.033 
\end{verbatim}}
The empirical distributions of the results show that the different randomizations did not affect the results qualitatively.

\paragraph{Average ranks for ties}

We also considered solution~(iii), that is, average ranks for ties. The p-values of the three tests of extreme-value dependence (applied in the same order as previously) were 0.6, 0.02 and 0, respectively. The p-values of the tests of exchangeability and goodness of fit were 0.12 and 0.18, respectively. The estimate of the parameter of the Gumbel--Hougaard copula was 1.446.

\paragraph{Random or average ranks for ties?}

The previous computations illustrate that solutions~(iii) and~(iv) for dealing with ties can result in significantly different conclusions. To gain insight into which solution should be preferred, if any, we designed an experiment tailored to the LOSS/ALAE data. Specifically, we simulated a large number of samples of size $n=1466$ from a Gumbel--Hougaard copula with parameter value 1.446, as suggested by the aforementioned parametric fit. We then modified each sample so that its marginal empirical c.d.f.s evaluated at the respective observations coincide with those of the LOSS/ALAE data. For instance, in the original data, the 27th to the 49th smallest values of LOSS are equal. Each simulated sample was modified so that the 27th to the 49th smallest values of the first variable all get replaced by the 49th smallest observation. The same approach was used for the second variable of the generated samples. Solutions~(iii) and~(iv) were applied next to each modified sample prior to running the tests of extreme-value dependence, and the resulting p-values were compared with those obtained by applying the tests on the corresponding unmodified sample (that is, with no ties). The code used to carry out the experiment for the test based on $S_{2n}$ defined in~\eqref{bk:eq:S2n} is given below:
{\small \begin{verbatim}
> mr.loss <- rank(myLoss[,1], ties.method="max")
> mr.alae <- rank(sort(myLoss[,2]), ties.method="max")
> test.func <- function(x) evTestK(x)$p.value
> do1 <- function()
+ {
+     x <- rCopula(1466, gumbelCopula(1.446))
+     y <- x[order(x[,1]),]
+     y[,1] <- y[mr.loss,1] 
+     y <- y[order(y[,2]),] 
+     y[,2] <- y[mr.alae,2] 
+     z <- apply(y, 2, rank, ties.method="random")
+     c(test.func(x), test.func(y), test.func(z))
+ }
> res <- t(replicate(1000, do1()))
> summary(round(res[,1] - res[,3],3))
     Min.   1st Qu.    Median      Mean   3rd Qu.      Max. 
-0.093000 -0.014000  0.001000  0.000033  0.014000  0.105000 
> summary(round(res[,1] - res[,2],3))
    Min.  1st Qu.   Median     Mean  3rd Qu.     Max. 
-0.54600 -0.25520  0.14000  0.06697  0.34750  0.52300 
> apply(res, 2, function(x) mean(x <= 0.05))
0.046 0.107 0.047 
\end{verbatim}}
For the test based on $S_{2n}$, the p-values computed from a continuous sample and the corresponding randomized sample are very close on average, the maximal deviation being relatively small. On the contrary, the p-values computed from a continuous sample are larger on average than the p-values computed from the corresponding sample involving average ranks, and the maximal deviation is very large. We also see that when solution~(iii) is considered, the test based on $S_{2n}$ is way too liberal, confirming the findings of \cite{bk:GenNesRup11}, while, when solution~(iv) is used, the test holds its level well. A similar experiment was performed for the test based on $T_{3,4,5,n}$ (with 100 replications only) and the conclusions are of the same nature but more pronounced:
{\small
\begin{verbatim}
> summary(round(res[,1] - res[,3],3))
    Min.  1st Qu.   Median     Mean  3rd Qu.     Max. 
-0.10400 -0.01425 -0.00150 -0.00050  0.01650  0.08700 
> summary(round(res[,1] - res[,2],3))
   Min. 1st Qu.  Median    Mean 3rd Qu.    Max. 
-0.0190  0.2700  0.4820  0.4536  0.6592  0.8730 
> apply(res, 2, function(x) mean(x <= 0.05))
0.05 0.45 0.05 
\end{verbatim}}

The previous experiment can be adapted to any data set containing ties and suggests that, in the case of the LOSS/ALAE data, solution~(iv) is meaningful while solution~(iii) should be avoided.

\section{Testing the maximum domain of attraction condition}

The statistical framework considered in the three previous sections can be regarded as the ``classical'' setting of dependence modeling by copulas. As mentioned in the introduction, modeling a copula by an extreme-value copula, or testing extreme-value dependence within such a framework, is particularly sensible if there are reasons to assume that the data at hand are generated by some maxima-forming process. If this is not the case, or if the hypothesis of extreme-value dependence is rejected, it might still be reasonable to make the (mild) assumption that the copula of interest lies in the domain of attraction of some extreme-value copula. It is the aim of the present section to briefly discuss how the latter assumption could be tested.

A precise formulation of the problem is as follows: we observe a sample of $d$-dimensional i.i.d.\ vectors
%(or, more generally, a stationary time series exhibiting weak serial dependence)
$\mathbf Y_1, \dots, \mathbf Y_n$ with c.d.f.\ $C^*\{G_1(y_1), \dots, G_d(y_d)\}$, where $G_1, \dots, G_d$ are assumed continuous and $C^*,G_1,\dots,G_d$ are unknown. We are interested in tests of
\begin{align} \label{bk:eq:evcondition}
H_0: C^* \in D(C) \text{ for some  } C \in \mathcal{EV}
\quad \text{against}\quad
H_1: C^* \notin D(C) \text{ for any } C \in \mathcal{EV},
\end{align}
where the notation $C^* \in D(C)$ is defined below~\eqref{bk:eq:maxdom}. Notice that the analogue univariate problem (i.e., testing the null hypothesis that the underlying distribution of a given univariate i.i.d.\ sample lies in the maximum domain of attraction of some extreme-value distribution) was tackled in \cite{bk:DieDehHus02}, \cite{bk:DreDehLi06} and \cite{bk:HusLi06}, while, in the multivariate case, only very few (validated) methods seem available.

A rejection of the null hypothesis in~\eqref{bk:eq:evcondition} gives indication that the stochastic dependence between componentwise block maxima formed from the $\mathbf Y_i$, no matter how large the blocks are, cannot be adequately described by an extreme-value copula. On the other hand, if the hypothesis is not rejected, it is promising to consider an extreme-value copula as a model provided the block size is sufficiently large. Also, in the latter case, one could make use of~\eqref{bk:eq:maxdom} to obtain the approximation that, for a sufficiently large~$r$, $C^*(\mathbf v) \approx \{ C(v_1^r, \dots, v_d^r) \}^{1/r}= C(\mathbf v)$ for all $\mathbf v \in [\mathbf t, \mathbf 1]$, with $\mathbf t =(t_1, \dots, t_d)$ close to $\mathbf 1$. This would imply that, at least in the upper tail, the copula $C^*$ can be well-approximated by an extreme-value copula $C$. A threshold model of that form was for instance considered in \cite{bk:LedTaw96} in a bivariate setting with generalized Pareto marginals.

A first promising approach to test $H_0$ in~\eqref{bk:eq:evcondition} consists of comparing two estimators of $C$ (or its characterizing objects) with different backgrounds. Under the null hypothesis, these estimators should not differ too much. Based on the peak-over-threshold method and in the bivariate case, \cite{bk:EinDehLi06} developed a test based on an Anderson--Darling-type statistic between two estimators of the stable-tail dependence function $\ell:[0,1]^2 \to \mathbb R$ defined by $\ell(x,y) = |x+y| A(x/|x+y|)$, where $A$ denotes the Pickands dependence function of $C$. Critical values for the test were obtained by approximately simulating from the limiting random variable. To the best of our knowledge, this testing procedure is the only validated method for testing the (bivariate) maximum domain of attraction condition.

A heuristic approach to test the null hypothesis in~\eqref{bk:eq:evcondition} in the bivariate case was described in \cite{bk:CorGenNes14}. Their method consists of considering a trimmed A-plot (see also Section~\ref{bk:subsec:pick}) defined by only including those points $(\hat T_i, \hat Z_i)$ in the set $\hat{\mathcal S}_n=\hat{\mathcal S}_n(\mathbf t)$ for which $\hat{\mathbf U}_i \in [\mathbf t, \mathbf 1]$, with some suitable thres\-hold parameter $\mathbf t =(t_1, t_2) \in [0,1]^2$ close to $\mathbf 1$. Based on the trimmed A-plot, the approach briefly described in Section~\ref{bk:subsec:pick} can be followed to obtain a B-spline smoothing estimator of the Pickands dependence function corresponding to the limiting extreme-value copula $C$. Plotting the residual sum of squared errors defined in~\eqref{bk:eq:resid} against the threshold $\mathbf t$ serves as a data-driven method for the choice of the threshold. For that particular choice, the A-plot as well as the testing procedure described in Section~\ref{bk:subsec:pick} can be used to assess heuristically whether the maximum domain of attraction condition holds or not.

Finally, the tests described in Section~\ref{bk:sec:tests} can be adapted to obtain simple heuristic procedures for testing$H_0$ in~\eqref{bk:eq:evcondition}. Under the null hypothesis, given $\mathbf Y_1, \dots, \mathbf Y_n$, if we form $k$ (componentwise) block maxima from blocks of length $m$,
\[
\mathbf X_{i} = (X_{i1}, \dots, X_{id}), \quad X_{ij}=  \max \{Y_{m(i-1)+1,j}, \dots, Y_{mi,j}\},
\]
$i \in \{1, \dots, k\}$, $j \in \{1, \dots d\}$,  where $km = n$ and $m$ is sufficiently large (if $n$ is not an integer multiple of $m$, then a negligible remainder block of length strictly smaller than $m$ occurs), then the copula of the block maxima $\mathbf X_i$ should (approximately) be an extreme-value copula. The tests described in Section~\ref{bk:sec:tests} could next be applied to $\mathbf X_1, \dots, \mathbf X_k$ to obtain an indication of whether the maximum domain of attraction condition holds or not. Another promising approach consists of adapting the approach in Section~\ref{bk:subsec:max} by only testing max-stability in the upper tail $[\mathbf t, \mathbf 1]$, with some suitable thres\-hold parameter $\mathbf t =(t_1, t_2) \in [0,1]^2$ close to $\mathbf 1$. This could be done by integrating the square of the process in~\eqref{bk:test_process} over the restricted set $[\mathbf t, \mathbf 1]$.

Precise asymptotic validations of these methods are, however, not available. A treatment of occurring bias terms from an undersized choice of the block length or the threshold parameter would be necessary, as for instance carried out in \cite{bk:BucSeg14} in an estimation framework for time series based on block maxima. Also, data-driven methods to choose the block length $m$ or the threshold parameter $\mathbf t$ would need to be developed.

%Finally, another simple alternative heuristic approach consists of combining the block maxima method and the tests described in Section~\ref{bk:sec:tests}. Under the null hypothesis, if we partition $\mathbf Y_1, \dots, \mathbf Y_n$ into $k$ (componentwise) block maxima of length $m$,
%\[
%\mathbf X_{i} = (X_{i1}, \dots, X_{id}), \quad X_{ij}=  \max \{Y_{m(i-1)+1,j}, \dots, Y_{mi,j}\},
%\]
%$i \in \{1, \dots, k\}$, $j \in \{1, \dots d\}$,  where $km = n$ and where $m$ is sufficiently large (if $n$ is not an integer multiple of $m$, then a negligible remainder block of length strictly smaller than $m$ occurs), then the copula of the block maxima $\mathbf X_i$ should be (approximately) given by an extreme-value copula.
%The tests described in Section~\ref{bk:sec:tests} could next be applied to $\mathbf X_1, \dots, \mathbf X_k$ to obtain an indication of whether the extreme-value condition holds or not. A precise asymptotic validation of this method is, however, not available. A treatment of occurring bias terms from an undersized choice of the block lengths would be necessary, as for instance carried out in an estimation framework for time series in \cite{bk:BucSeg14}. Also, data-driven methods to choose $m$ and $k$ would need to be developed.

\section{Open questions and ignored difficulties}

Several issues dealt with in this chapter would need to be thoroughly investigated in future research. For instance, the suggested approach for handling ties in data sets for which it is actually reasonable to assume that the apparent discontinuities are only due to precision or rounding issues would need to be studied more in depth. While for the LOSS/ALAE data set, it seemed reasonable to break ties at random a large number of times, this may not be the case for other data sets in which the proportion of ties is significantly larger \citep[see e.g.][Section~4]{bk:GenNesRup11}. Yet, even more difficult appears to be the problem of testing extreme-value dependence from truly discontinuous observations such as count data. A promising starting point for adapting some of the statistical procedures described in this work to such a context is the recent work of \cite{bk:GenNesRem14} on the {\em multilinear empirical copula}.

With financial applications in mind in particular, tests of extreme-value dependence would also need to be adapted to multivariate stationary times series. The methods briefly described in this chapter all rely on the assumption that the observations at hand are serially independent which is hardly verified for many data sets of interest. Applying the discussed statistical procedures to (almost i.i.d.) standardized residuals from common time series models might be an option, but it is unclear whether the necessary additional estimation step affects the limiting null distribution of the test statistics or not. For that purpose, a starting point might be the work of \cite{bk:Rem10} where the asymptotics of the empirical copula process of standardized residuals are investigated. If the tests are to be applied on the stationary raw time series data, then their empirical levels will most likely be affected by the serial dependence present in the observations. In such a situation, the dependent multiplier bootstrap studied in \cite{bk:BucKoj14} could be used to adapt some of the reviewed tests of extreme-value dependence.

\smallskip

\textbf{Acknowledgments} The authors are grateful to Christian Genest, Johanna Ne\v slehov\'a and an anonymous referee for their constructive comments on an earlier version of this chapter. This work has been supported in part by the Collaborative Research Center \textit{Statistical Modeling of Nonlinear Dynamic Processes} (SFB 823, project A7) of the German Research Foundation (DFG).

\bibliographystyle{asa}
\bibliography{biblio}

\end{document}